\nopagenumbers 
\magnification=\magstep1  
\null  
\centerline{ \bf RAPIDITY GAPS IN MINIJETS PRODUCTION} 
\centerline{ \bf AT TEVATRON ENERGIES} 
\vskip .25in 
\centerline{G. Calucci, R. Ragazzon and D. Treleani} 
\vskip .05in 
\centerline{\it Dipartimento di Fisica Teorica dell'Universit\`a  
and INFN Sezione di Trieste} 
\centerline{\it Trieste, I 34014 Italy} 
\vskip .25in
\centerline{ABSTRACT} 
\vskip .25in 
\midinsert 
\narrower\narrower  
\noindent   
Multiparton interactions modify the high energy hadronic cross section to
produce minijets with a rapidity gap in the distribution of
secondaries. At Tevatron
energy the correction to the single scattering term is 
large for transverse momenta smaller than $6GeV$.
\endinsert  
\vskip .25in
\par 	One of the topics of interest in perturbative QCD is to test
the BFKL approach to semihard hadron interactions. The BFKL approach
allows one to write an explicit expression for the inclusive cross section to 
produce minijets as a convolution of the partonic
cross section with the effective parton structure functions
(gluon plus 4/9 quark and anti-quark structure functions) [1]:

$${d\sigma_L\over dx_Adx_Bd^2k_ad^2k_b}= 
  f_{eff}(x_A,k_a^2)f_{eff}(x_B,k_b^2) 
  {d\hat{\sigma}_L\over d^2k_ad^2k_b}\eqno(1)$$ 

\par\noindent
where $x_A$, $x_B$ are momentum fractions of the interacting
partons and $k_a$, $k_b$ are the transverse momenta of the two observed
minijets. The partonic cross section is written as
 
$${d\hat{\sigma}_L\over d^2k_ad^2k_b}=\biggl[{N_c\alpha_s\over 
k_a^2}\biggr] 
     f(k_a,k_b,\Delta y)\biggl[{N_c\alpha_s\over k_b^2}\biggr]\eqno(2)$$ 
 
\par\noindent 
where $N_c$ is the number of colors, $\alpha_s$ is the strong 
coupling 
constant and $f(k_a,k_b,\Delta y)$ is the solution to the BFKL equation depending
on the rapidity distance $\Delta y$ of the two observed minijets. The partonic cross section
in Eq.(2) exibits Regge behaviour at large values of $\Delta y$. Actually 

$$\hat{\sigma}_L\to e^{(\alpha_P-1)\Delta y}\eqno(3)$$

\par\noindent
and the Regge intercept $\alpha_P$ is computed in perturbative QCD. 
In fact the inclusive
parton cross section in Eq.(2) corresponds to a parton interaction with a
cut BFKL Pomeron exchange. In analogy with soft processes one may
therefore consider the possibility to produce the two observed minijets
without cutting the BFKL Pomeron, namely by an elastic parton-parton
collision represented with an uncut BFKL Pomeron exchange[2]. The 
singlet contribution to elastic
parton process is related to the inelastic one by the unitarity relation.
The behaviour at large $\Delta y$ is therefore:

$$\hat{\sigma}_{S}\to e^{2(\alpha_P-1)\Delta y}\eqno(4)$$

\par\noindent
and the corresponding experimental signature is a rather spectacular one.
Two minijets separated by a large gap in the production of secondaries.
Indeed to compare the rates of the two parton processes, inelastic and
elastic, one needs to make the further request that soft interactions between
spectator partons in the hadronic event are not going to fill
the rapidity gap. The elastic partonic cross section needs therefore to be
multiplyed by the survival probability factor $\langle S^2\rangle$[3], which
however is expected to depend smootly on the rapidity interval $\Delta y$ 
in such a way that it would provide only an overall rescaling factor to the elastic 
partonic cross section.
\par
The point which we like to focus on is unitarization. In fact the relation 
between elastic and inelastic parton cross section is the unitarity relation
applied to the parton-parton collision. If one looks rather to the the problem
of unitarizing the whole semihard hadronic process one realizes that in the
semihard region unitarity induces further sources of minijets other than
the cut and the uncut BFKL Pomerons already considered. In fact unitarity
introduces in the hadronic interaction also multiple parton collsions whose
effect is to change the simple relation between semihard events with
and without rapidity gap discussed above.
\par\noindent
The argument for multiple parton collisions is the following. One may integrate 
the inclusive cross section for production of minijets, Eq.(1), with the cutoff
$q_t^{min}$ which represents the lower treshold for observing a parton as 
a minijet in the final state. At large energy in the hadron-hadron c.m. system
the integrated cross section may easily be larger with respect to the total 
inelastic cross section when $q_t^{min}$ is relatively small[4].
On the other hand the partonic cross section 
is in comparison still rather small. The large value  
of the integrated inclusive cross section is therefore the consequence 
of the large flux of partons in the initial state, which may give rise 
to an average number of partonic collisions larger than one,
explaining in this way the large value of the integrated inclusive 
cross section[5]. 
\par
Moving $q_t^{min}$ towards smaller values one is therefore entering the region
where multiple parton collisions are an important effect. If one continues to
decrease the value of $q_t^{min}$ one encounters a further problem.
One expects in fact that the value of $\alpha_S$ which one 
should  use will increase at smaller values of $q_t^{min}$. The BFKL expressions 
for the inelastic and elastic parton cross sections grow as the 
exponential of  
$\alpha_S$ and on $2\alpha_S$ 
respectively, in such a way that the rise of the elastic 
parton cross section is too fast with respect to the rise of the inelastic one.
\par\noindent
To have an indication on the values of $q_t^{min}$ where the
problem may occur we have taken the simplest attitude.
In analogy to the $s$-channel unitarization of the soft Pomeron exchange
we have included in the semihard partonic interaction the
exchange of two BFKL Pomerons and we have used the AGK 
cutting rules[6] to obtain the inelastic contributions to the  
cross section. The semihard partonic cross section  
$\hat{\sigma}_H(y)$ is therefore expressed as      
 
$$\hat{\sigma}_H(y)=\hat{\sigma}_{S}(y)+ 
  \bigl(\hat{\sigma}_L(y)-4\hat{\sigma}_{S}(y)\bigr)+ 
  2\hat{\sigma}_{S}(y) 
  \eqno(5)$$ 
  
\par\noindent 
where the `elastic' partonic cross  
section is identified with the `diffractive' cut of the double BFKL 
Pomeron  
exchange contribution to the forward parton amplitude[7]. 
The single BFKL Pomeron exchange contributes with 
$\hat{\sigma}_L(y)$ and the contributions from the double 
BFKL Pomeron exchange, according with the AGK cutting rules,  
are: $\hat{\sigma}_{S}(y)$, the `diffractive' 
contribution, $-4\hat{\sigma}_{S}(y)$, the one Pomeron cut, and 
$+2\hat{\sigma}_{S}(y)$, the two Pomeron cut. Eq.(5) allows  
one to define the kinematical region of applicability of the approach. 
Since the one BFKL cut Pomeron contribution 
to the cross section must be positive: 
 
$$\bigl(\hat{\sigma}_L(y)-4\hat{\sigma}_{S}(y)\bigr)>0\eqno(6)$$ 
 
\par\noindent 
If one 
uses for $\alpha_S$ the expression of the running coupling constant 
one may obtain, from the bound in Eq.(6), a relation between $q_t^{min}$
and the average rapidity interval of two interacting partons, which is
easily translated into a relation between $q_t^{min}$ and the hadron-hadron
c.m. energy. When moving towards smaller values of $q_t^{min}$ 
one can therefore distinguish three different regimes: 
\item{I-} The cutoff is sizeable with respect to the typical energy  
available to the semihard partonic interaction. The corresponding  
`elementary' parton interaction  
is small, no unitarization is needed and the semihard cross section  
is well described by a single partonic collision. 
\item{II-} At relatively smaller values of the cutoff
a single partonic interaction is 
still well described by the BFKL dynamics. The semihard 
hadronic cross section is however too large with respect to the 
total inelastic cross section and unitarity corrections are to be  
taken into account. The unitarization of the hadronic semihard  
cross section is achieved by taking into account multiparton 
interactions, namely different pairs of partons  
interacting independently with BFKL Pomeron exchange. 
Typically the different partonic interactions are localized at different  
points in the transverse plane, in the region of overlap  
of the matter distribution of the two hadrons. 
\item{III-} With even smaller values of the cutoff one may still be  
in the regime where perturbative QCD can be used, since the value of 
$\alpha_S(q_t^{min})$ is small, but the elastic parton cross section is too
large with respect to the inelastic one and the `elementary' 
parton process is not well described any more by the single 
BFKL Pomeron exchange.
One may obtain an indication on the limits 
between  
regions II and III by testing whether the bound in Eq.(6) is 
satisfied. 
\par
We focus our attention on region II, where one expects to find non
trivial effects from unitarization while the perturbative part of the
semihard interaction is described within the BFKL approach. Although
the perturbative part of the interaction is explicitly
known also in region II, one is not yet in the
position to write explicitly the semihard cross section. The reason
is that the non perturbative input in this case is rapresented by 
the multiparton distributions[8], which are a piece of information on the
hadron structure independent of the hadron structure functions
of large $p_t$ physics. In fact multiparton distributions are dimensional
quantities and have an explicit dependence on the multiparton correlations. The usually
considered parton structure
functions on the contrary can carry information only on the average number 
of partons. At present the only information on the multiparton
distribution is an indication on the scale factor which gives the dimensionality to the 
multiparton distributions[9]. We take therefore the simplest attitude
namely we consider the simplest case where multiparton correlations 
are neglected and we express the multiparton distributions 
as Poissonians with a scale factor consistent with the 
experimental indication[9].
If one in addition neglects the possibility
of having semihard parton rescatterings in the semihard interaction, 
one obtains for the semihard
hadronic cross section $\sigma_H$ the simple eikonal form:  
 
$$\sigma_H=\int d^2\beta\Bigl[1-{\rm exp}\bigl(-\Phi(\beta)\bigr)\Bigr] 
=\int d^2\beta\sum_{\nu=1}^{\infty}{\bigl[\Phi(\beta)\bigr]^{\nu}\over\nu!}
{\rm exp}\bigl(-\Phi(\beta)\bigr)
  \eqno(7)$$ 
 
\par\noindent 
where 
 
$$\Phi(\beta)\equiv\Phi_{S}(\beta)+\Phi_P(\beta)\equiv 
  \int_{y_m}^{y_M}dy\int_{y}^{y_M}dy' 
  \bigl(\phi_{S}(\beta;y,y')+\phi_P(\beta;y,y')\bigr)\eqno(8)$$ 
 
\par\noindent 
with  
 
$$\phi_{S,P}(\beta;y,y')\equiv\int d^2bD_A\bigl(b,x(y)\bigr) 
  \hat{\sigma}_{S,P}(y'-y)D_B\bigl(b-\beta,x'(y')\bigr) 
  \eqno(9)$$ 
 
\par\noindent 
and $y_M$, $y_m$ are the maximum and minimum rapidity values 
allowed by kinematics. $D\bigl(b,x(y)\bigr)$ is the average number
of partons with transverse coordinate $b$ and fractional momentum $x$,
which may be expressed as a function of the rapidity $y$ of the produced
minijet. The indices $A$ and $B$ refer to the two interacting hadrons.
\par	One is interested in the component of $\sigma_H$ 
which represents two minijets at rapidities 
$\bar{y}$ and $\bar{y}'$, in the central rapidity region, 
with associated gap $\Delta y=\bar{y}'-\bar{y}$  
in the rapidity distribution of secondary produced gluons. 
To that purpose one needs to exclude in Eq.(7) both the elastic terms, 
with final state minijets in the gap, and all the inelastic 
partonic interactions, generated with elementary probability 
$\phi_P$. 
The cross section to observe two minijets at rapidities 
$\bar{y}$ and $\bar{y}'$, with the gap $\Delta y=\bar{y}'-\bar{y}$  
in the rapidity distribution 
of secondaries, is therefore expressed as 
 
$$\eqalign{ 
  {d\sigma_H(\Delta y)\over d\bar{y}d\bar{y}'}=\int d^2\beta 
  \Biggl[\sum_{\nu=1}^\infty\nu  
  \phi_{S}(\beta;\bar{y},\bar{y}') 
  {\bigl[\Phi_{S}(\beta,\Delta y)\bigr]^{\nu-1}\over\nu!}&\cr 
 +\sum_{\nu=2}^\infty\nu(\nu-1)  
  \int_{y_m}^{\bar{y}}dy\phi_{S}(\beta;y,\bar{y}') 
    \int_{\bar{y}'}^{y_M}&dy'\phi_{S}(\beta;\bar{y},y')\cr 
  \times&{\bigl[\Phi_{S}(\beta,\Delta y)\bigr]^{\nu-2}\over\nu!} 
  \Biggr]e^{-\Phi(\beta)}} 
  \eqno(11)$$ 
 
\par\noindent 
After summing on $\nu$ one obtains 
 
$$\eqalign{ 
  {d\sigma_H(\Delta y)\over d\bar{y}d\bar{y}'}=\int d^2\beta 
  \Bigl[\phi_{S}(\beta;\bar{y},\bar{y}') 
  +\int_{y_m}^{\bar{y}}&dy\phi_{S}(\beta;y,\bar{y}') 
    \int_{\bar{y}'}^{y_M}dy'\phi_{S}(\beta;\bar{y},y')\Bigr]\cr 
  \times&{\rm exp}\Bigl\{ 
   \Phi_{S}(\beta;\Delta y)-\Phi_{S}(\beta)-\Phi_P(\beta)\Bigr\} 
  }\eqno(12)$$ 
 
\par 
The two addenda in Eq.(12) are the single and double `elastic' 
scattering contributions.  
In the single scattering term both observed minijets are 
produced in the same elementary partonic interaction, in the 
double scattering term the two minijets are generated in different partonic 
collisions. Both terms are multiplied by the absorption factor 
$exp\bigl\{-\bigl(\Phi_S(\beta)-\Phi_S(\beta;\Delta y)\bigr)\bigr\}$  
that removes the `elastic' parton interactions which would fill the gap,  
actually those which produce minijets with rapidities $y$ and $y'$ 
such that ${\bar y}\le y$ or $y'\le\bar{y}'$. 
At a fixed value of $\beta$ the cross section is multiplied 
by $exp\bigl\{-\Phi_P(\beta)\bigr\}$ which is the probability of not
having any inelastic partonic interaction in the process.
One may recognize in Eq.(12) the semihard contribution to the
survival probability factor $\langle S^2(\beta)\rangle$ of ref.[3]. Actually 
$exp\bigl\{-\Phi_S(\beta)-\Phi_P(\beta)\bigr\}$ is the
probability factor of not having any semihard activity in the
spectator partons.
\par 
To have a quantitative indication on the boundaries  
of the kinematical regions,  
we have worked out a numerical example. 
The input is the average number of partons $D(b,x)$ and the `elementary'  
partonic cross sections $\hat{\sigma}_{L,S}$. We have factorized $D(b,x)$ as 
$f_{eff}(x)\times F(b)$, where $f_{eff}(x)$ is the effective structure 
function and $F(b)$ is a gaussian, normalized 
to one and such as to give for the double scattering term a scale factor 
consistent with the experimental indication of $\sigma_{eff}$[9].  
In our numerical example we have chosen $\sigma_{eff}=20mb$ and 
as a scale factor for the structure functions  
we have taken $q_t^{min}/2$. $\alpha_S$ is a free parameter in the BFKL 
approach, it is not a running coupling constant, 
one expects however that the value of $\alpha_S$ which one 
should use is not too different from the value of 
the running $\alpha_S$ at the scale of the typical momentum transferred 
in the process. We have chosen as a value of $\alpha_S$ the value 
of the running coupling computed with $q_t^{min}/2$ as a scale factor. 
The values of the semihard cross section $\sigma_H$, as expressed in Eq.(7),
which we obtain with this input are consistent with the experimental values
published by UA1[10]. 
\par	An estimate of the boundaries between the regions I, II and III 
discussed above is shown in fig.1. The boundary between region I and II
is obtained by requiring that the unitarized expression for $\sigma_H$
is $20\%$ smaller with respect to the single scattering term. The boundary
between regions II and III is obtained by saturating on the average,
namely after integrating with the structure functions, the bound in
Eq.(6). The effect on the cross section to produce minijets with rapidity
gap is shown in fig.2, where we plot the 
cross section in Eq.(12) divided by the survival probability factor
$exp\{-\Phi_S(\beta)-\Phi_P(\beta)\}$. The process is $p\bar{p}$
at $\sqrt{s}=1.8TeV$ and $q_t^{min}=5GeV$. The continuous
curve is obtained by using the value $\sigma_{eff}=20mb$ as a input
and the dashed curve is obtained with the value $\sigma_{eff}=12mb$.
The dotted curve is the result of the single scattering term alone.   
\par 
As it is shown in fig.2
the effect of unitarization on the behaviour of the 
cross section is large. 
In the actual case the main modification to the dependence on $\Delta y$
is due to the presence of multiple elastic parton scatterings whose effect
on the cross section is twofold. A different dependence on $\Delta y$, with
respect to the single scattering term, is
induced by the contribution of the process where the two observed minijets 
originate in different elastic partonic interactions, 
the second term in Eq.(12).
A second source for the different dependence on $\Delta y$ is the correction
induced by multiple elastic scatterings to the survival probability factor. In fact
not all undelying hadron activity needs to be excluded. 
Elastic parton scatterings
which produce minijets outside the gap are allowed and the corresponding 
contribution to the cross section depends on $\Delta y$.
The effect of the inelastic semihard 
partonic interactions is, on the contrary, factorized 
at fixed impact parameter $\beta$ and independent on $\Delta y$.
The main effect of the inelastic partonic processes is to contribute  
to the survival probability $\langle S^2\rangle$ of ref.[3] 
rather than modifying the dependence on $\Delta y$. 

\vskip .25in

\par    {\bf References} 
\vskip .15in 
\item{1.} A.H. Mueller and H. Navelet, {\it Nucl. Phys.}
{\bf B282}, 727 (1987);
V. Del Duca, preprint DESY 95-023, DFTT 13/95 (unpublished).
\item{2.} A.H. Mueller and W.-K. Tang, {\it Phys. Lett.} {\bf B284}, 123 
(1992); V. Del Duca and W.-K. Tang, {\it Phys. Lett.} {\bf B312}, 225 
(1993).
\item{3.} J.D. Bjorken, {\it Phys. Rev.} {\bf D47}, 101 (1992). 
\item{4.} G. Pancheri and Y. Srivastava {\it Phys. Lett.} {\bf B182}, 
199 (1986); S. Lomatch, F.I. Olness and J.C. Collins {\it Nucl. Phys.} 
{\bf B317}, 617 (1989). 
\item{5.} M. Jacob and P.V. Landshoff, {\it Mod. Phys. Lett.} {\bf A1}, 
657 (1986); 
Ll. Ametller and D. Treleani, {\it Int. J. Mod. Phys.} {\bf A3}, 521 
(1988). 
\item{6.} V. Abramovskii, V.N. Gribov and O.V. Kancheli, {\it Yad. Fiz.} 
{\bf 18}, 595 (1973) [{\it Sov. J. Nucl. Phys.} {\bf 18}, 308 (1974) ]. 
\item{7.} L.V. Gribov, E.M. Levin and M.G. Ryskin, {\it Phys. Rep.} 
{\bf 100}, 1 (1983). 
\item{8.} H.D. Politzer, {\it Nucl. Phys.} {\bf B172}, 349 (1980); 
R.K. Ellis, R. Petronzio and W. Furmanski, {\it ibid.} {\bf B207},  
1 (1981); 
N. Paver and D. Treleani, {\it Nuovo Cimento} {\bf A70}, 
215 (1982); Zeit.  
Phys.{\bf C28}, 187 (1985); B. Humpert, {\it Phys. Lett.}  
{\bf 131B}, 461 (1983); 
B. Humpert and R. Odorico, {\it ibid} {\bf 154B}, 211 (1985);  
T. Sjostrand and 
M. Van Zijl, {\it Phys. Rev.} {\bf D36}, 2019 (1987).
\item{9.} F. Abe et al., {\it Phys. Rev.} {\bf D47}, 4857 (1993). 
\item{10.} C. Albajar et al. {\it Nucl. Phys.} {\bf B309}, 
405 (1988).

\vfill 
\eject 
 
\par    {\bf Figure captions} 
\vskip .15in 
\item{Fig.1-} The three different kinematical regions which characterize 
semihard hadronic interactions. I: only the single partonic collision, 
described by a single BFKL Pomeron exchange, 
is relevant; II: multiparton collisions are to be taken into 
account, each partonic 
interaction is however well described by single BFKL Pomeron exchange; 
III: the single BFKL Pomeron exchange is not any more an adequate 
description of the single parton interaction. 
\vskip .15in 
\item{Fig.2-} Cross section for production of minijets  
with rapidity gap as a function of the gap $\Delta y$. The process
is $p\bar{p}$ at $\sqrt{s}=1.8TeV$ and $q_t^{min}=5GeV$. 
The dotted curve is the 
single scattering term contribution without unitarity
corrections; the continuous and dashed curves include all 
disconnected multiple parton collisions. The continuous curve
is obtained with $\sigma_{eff}=20mb$ and the dashed curve
with $\sigma_{eff}=12mb$.
\vfill 
\eject\end  
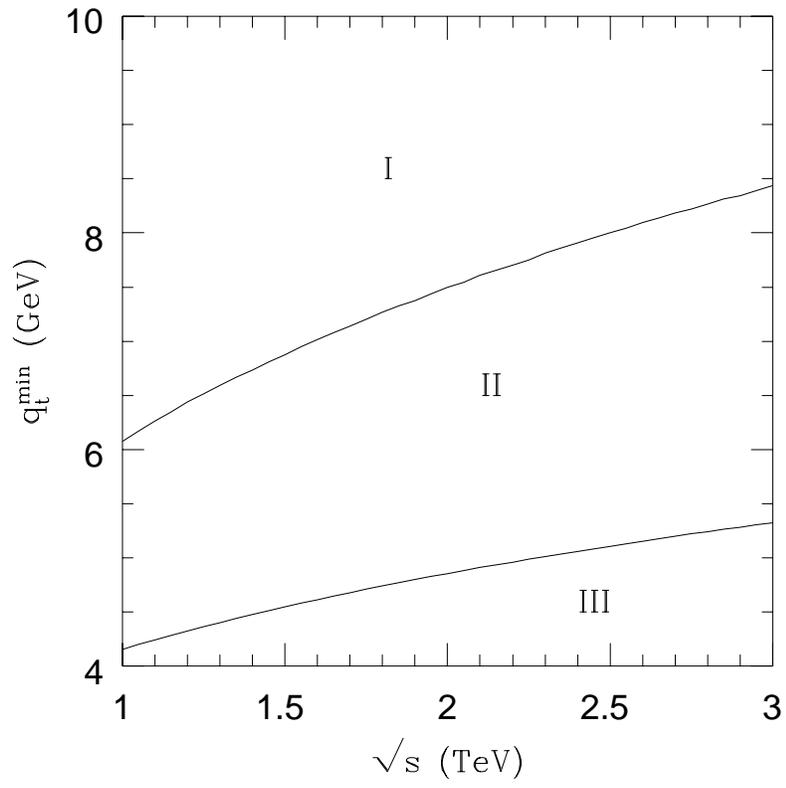
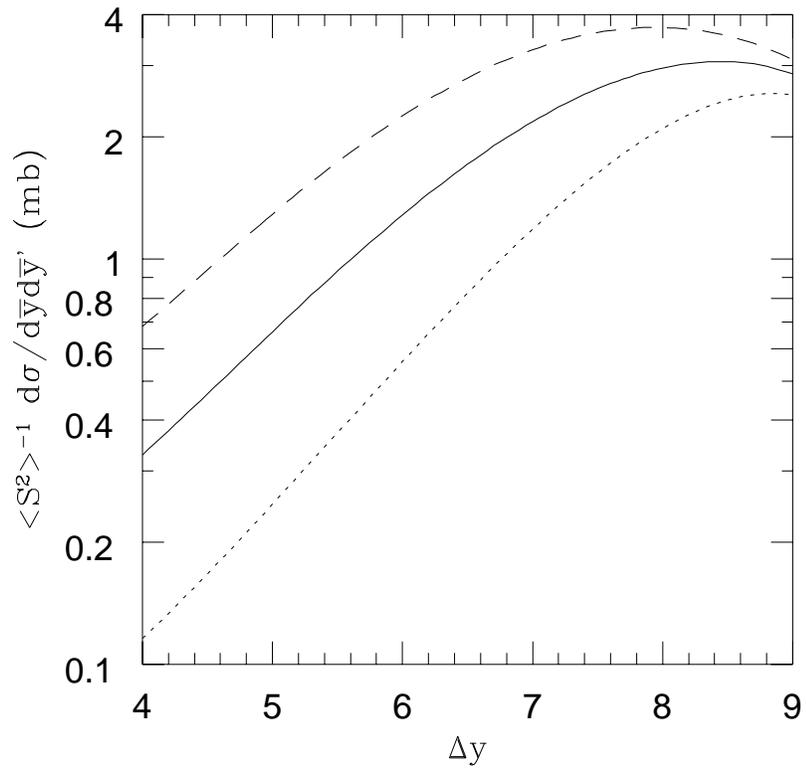
\bye